\documentclass{iopart}
\usepackage{iopams,amsfonts,amssymb,amsthm} 
\usepackage{geometry}                
\usepackage{graphicx}
\usepackage{rawfonts}
\usepackage{amssymb}
\usepackage{epstopdf}
\usepackage[usenames,dvipsnames,svgnames]{xcolor}
\input prepictex.tex
\input pictex.tex
\input postpictex.tex

\newtheorem{theo}{Theorem}
\newtheorem{lemm}{Lemma}

\def\Pr{\noindent \emph{Proof: }}
\def\qed{$\Box$}

\def\nor{\normalsize}
\def\fns{\footnotesize}

\def\sfrac#1#2{\hbox{\nor $\frac{#1}{#2}$}}
\def\Sfrac#1#2{\hbox{\large $\frac{#1}{#2}$}}

\def\Ref#1{(\ref{#1})}

\def\L{\left(} \def\R{\right)}

\begin{document}

\title{Force-induced desorption of $3$-star polymers: a self-avoiding walk model}
\author{EJ Janse van Rensburg$^*$ and S G  Whittington$^\dagger$ }
\address{
{}$^*$Department of Mathematics \& Statistics, York University, M3J 1P3, Toronto, Canada \\
{}$^\dagger$Department of Chemistry, University of Toronto, M5S 3H6, Toronto, Canada 
}

\begin{abstract}
We consider a simple cubic lattice self-avoiding walk model of 3-star polymers adsorbed at
a surface and then desorbed by pulling with an externally applied force. We determine
rigorously the free
energy of the model in terms of properties of a self-avoiding walk, and 
show that the phase diagram includes 4 phases, namely a ballistic phase
where the extension normal to the surface is linear in the length,
an adsorbed phase and a mixed phase, in addition to the free phase where the model is neither
adsorbed nor ballistic. In the adsorbed phase all three branches or arms of the star are adsorbed
at the surface. In the ballistic phase two arms of the star are pulled into a ballistic phase, while
the remaining arm is in a free phase. In the mixed phase two arms in the star are adsorbed while
the third arm is ballistic. The phase boundaries separating the ballistic and mixed phases, and
the adsorbed and mixed phases, are both first order phase transitions.  The presence of the 
mixed phase is interesting because it doesn't occur for pulled, adsorbed self-avoiding walks.  In an 
atomic force microscopy experiment it would appear as an additional phase transition as a function of force.

\end{abstract}

\maketitle

\section{Introduction}
\label{sec:Introduction}
The standard model of linear polymers in dilute solution in a good solvent is 
self-avoiding walks \cite{Hammersley1957,Rensburg2015,MadrasSlade}.  If the 
walk is attached to a surface at which it can adsorb then we know that the 
model has an adsorption transition where the free energy is non-analytic.  
There is a critical temperature below which the walk is adsorbed at the surface 
and the fraction of vertices in the surface is non-zero.  Above this temperature 
the walk is desorbed \cite{HTW}.  For numerical estimates of the location of the 
transition see for instance \cite{Guttmann2014,RensburgRechnitzer} and references therein.

In the last few years there has been considerable interest in 
how self-avoiding walks respond to a force 
\cite{Beaton2015,GuttmannLawler,IoffeVelenik,IoffeVelenik2010,Rensburg2009,Rensburg2016a},
for instance as a model of polymers subjected to forces 
in atomic force microscopy \cite{Haupt1999,Zhang2003}.  
For a review see \cite{Orlandini}.  A particularly interesting situation is when 
the walk is adsorbed at a surface and is then desorbed by a force applied 
at the last vertex of the walk
\cite{Guttmann2014,Rensburg2013,Krawczyk2005,Krawczyk2004,Mishra2005}.  
For related work see for instance \cite{Skvortsov2009} and \cite{Binder2012}.
There are interesting similarities and differences when the force is applied in different ways
\cite{Rensburg2016b,Rensburg2017}.  In the next section we give a brief review of the 
available results.

\begin{figure}[t]
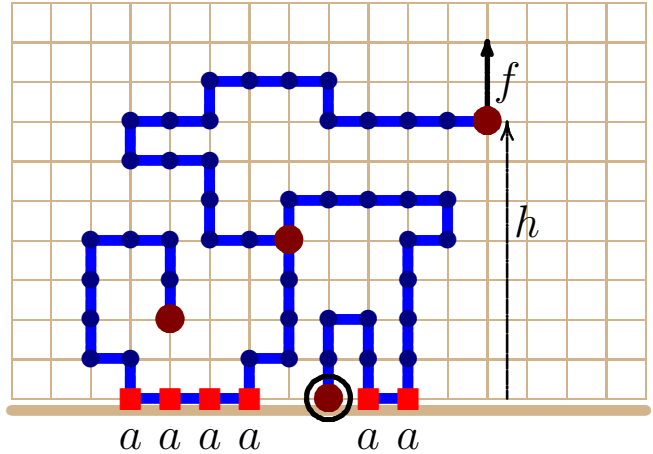

\beginpicture
\setcoordinatesystem units <1.5pt,1.5pt>
\setplotarea x from -140 to 110, y from -10 to 100
\setplotarea x from -80 to 80, y from 0 to 100

\color{Tan}
\grid 16 10 
\color{Tan}
\setplotsymbol ({$\bullet$})
\plot -80 -3 80 -3 /

\color{Blue}
\plot 0 0 0 10 0 20 10 20 10 10 10 0 20 0 20 10 20 20 20 30 20 40 30 40 30 50 20 50 
10 50 0 50 -10 50 -10 40   /
\plot -10 40 -20 40 -30 40 -30 50 -30 60 -40 60 -50 60 -50 70 -40 70 -30 70 -30 80 
-20 80 -10 80 0 80 0 70 10 70 20 70 30 70 40 70 /
\plot -10 40 -10 30 -10 20 -10 10 -20 10 -20 0 -30 0 -40 0 -50 0 -50 10 -60 10
-60 20 -60 30 -60 40 -50 40 -40 40 -40 30 -40 20  /
\color{NavyBlue}
\multiput {\LARGE$\bullet$} at 
0 0 0 10 0 20 10 20 10 10 10 0 20 0 20 10 20 20 20 30 20 40 30 40 30 50 20 50 
10 50 0 50 -10 50 -10 40  -20 40 -30 40 -30 50 -30 60 -40 60 -50 60 -50 70 
-40 70 -30 70 -30 80 -20 80 -10 80 0 80 0 70 10 70 20 70 30 70 -10 30 
-10 20 -10 10 -20 10 -20 0 -30 0 -40 0 -50 0 -50 10 -60 10
-60 20 -60 30 -60 40 -50 40 -40 40 -40 30 /

\color{red}
\multiput {\large$\blacksquare$} at 10 0 20 0 -20 0 -30 0 -40 0 -50 0 /

\setplotsymbol ({\LARGE$\cdot$})
\color{black}
\arrow <5pt> [.2,.67] from 40 70 to 40 90
\put {\LARGE$f$} at 45 80
\multiput {\LARGE$a$} at 10 -10 20 -10 -20 -10 -30 -10 -40 -10 -50 -10 /
\circulararc 360 degrees from 5.5 0 center at 0 0 

\color{Maroon}
\put {\huge$\bullet$} at 0 0 
\circulararc 360 degrees from 0 3 center at 0 0
\put {\huge$\bullet$} at -10 40 
\circulararc 360 degrees from -7 40 center at -10 40 
\put {\huge$\bullet$} at 40 70  
\circulararc 360 degrees from 43 70 center at 40 70  
\put {\huge$\bullet$} at -40 20  
\circulararc 360 degrees from -37 20 center at -40 20  

\color{black} \normalcolor

\setplotsymbol ({\fns$\cdot$})
\color{black}
\arrow <7pt> [.2,.67] from 45 0 to 45 70
\put {\LARGE$h$} at 50 45

\endpicture

\caption{A $3$-star in the square lattice.  The star is fixed at the origin in 
a vertex of degree $1$, and pulled by a vertical force $f$ at another
vertex of degree $1$.  Vertices in the star bind to the horizontal 
adsorbing line with activity or weight $a$.  These are visits of the
star to the adsorbing line and there are $v=6$ visits in this case. 
The height of the pulled vertex is denoted
by $h$ and in this case $h=7$. }
\label{fig:sketch}  
\end{figure}

A natural question that arises is how the architecture of a polymer affects its properties.
This was investigated for lattice polygons (a model of ring polymers) 
\cite{Guttmann2017} and we have a fairly complete understanding of the behaviour 
in three dimensions.  Beaton \cite{Beaton2017} has given an essentially 
complete solution for staircase polygons in 
two dimensions.  In this paper we look at the problem of uniform $3$-star polymers 
in $\mathbb{Z}^3$ desorbed from a surface by an applied force and with a vertex
of degree $1$ attached to an adsorbing $2$-dimensional lattice plane, 
and pulled at another vertex of degree $1$. This is schematically illustrated in 
figure \ref{fig:sketch}.

A \emph{star} with $f$ arms, or an \emph{$f$-arm star}, is a connected
graph with no cycles, one vertex of degree $f$ and $f$ vertices of degree $1$.  A
star with $f$ arms is also called an $f$-star and it is embedded in $\mathbb{Z}^d$
if each arm is a self-avoiding walk and if the arms are mutually avoiding 
in the lattice.  In this case it is a \emph{lattice star} (or, in this paper, a \emph{star}).
An arm is also called a \emph{branch} and we use the terms interchangeably.  A lattice 
star is \emph{uniform} if all arms have the same number 
of edges.  We shall count embeddings in the hypercubic lattice, $\mathbb{Z}^d$,
of lattice stars with one end-vertex of degree $1$ fixed at the origin.
Write $s_n^{(f)}$ for the number of such embeddings 
with a total of $n$ edges.  Note that $f$ must divide $n$.  For the $d$-dimensional 
hypercubic lattice $\mathbb{Z}^d$ with $f=3, \ldots, 2d$, we know that 
\cite{WhittingtonSoteros1992}
\begin{equation}
\lim_{n\to\infty} \sfrac{1}{n} \log s_n^{(f)} = \log \mu_d
\label{eqn:stargrowth}
\end{equation}
where $\mu_d$ is the growth constant of the $d$-dimensional 
hypercubic lattice (and the limit $n\to\infty$ is taken through $n=fm$
(multiples of $f$ in $\mathbb{N}$)).  
The growth constant is defined precisely in Section 2.
We shall usually omit the 
words \emph{uniform} and \emph{lattice} since we shall be concerned almost exclusively
with uniform stars embedded in $\mathbb{Z}^d$.

Denote by $(x_1,x_2,x_3)$ the coordinates of a vertex in $\mathbb{Z}^3$.
All the vertices have integer coordinates.
The \emph{adsorbing plane} in $\mathbb{Z}^3$ is the plane $x_3=0$.
We are primarily concerned with $3$-stars with a vertex of degree $1$
in the adsorbing plane (attached at the origin).  The adsorbing plane divides 
the lattice and the star is confined to the upper half-lattice where it is pulled
from another vertex of degree $1$ by a vertical force $f$ (in the $x_3$-direction).  

If the star has $v+1$ vertices in $x_3=0$ (these are \emph{visits} in the
adsorbing plane), then it is weighted by a factor $a^v$.  The \emph{height}
of the vertex where the star is pulled by a force is denoted by $h$ and 
the star will be weighted by a factor $y^h$.  We shall write $s_n^{(f)}(v,h)$ 
for the number of $f$-stars with these conditions.  Define the 
partition function
\begin{equation}
S_n^{(f)}(a,y) = \sum_{v,h} s_n^{(f)}(v,h) \, a^v y^h.
\label{eqn:starPF}
\end{equation}
Our aim is to determine the free energy
\begin{equation}
\sigma^{(f)}(a,y) = \lim_{n\to\infty}\sfrac{1}{n} \log S_n^{(f)}(a,y)
\label{eqn:starFE}
\end{equation}
in terms of properties of self-avoiding walks, and hence determine
features of the phase diagram in the $(a,y)$-plane.  We do this, in particular,
for \emph{adsorbing and pulled} $3$-stars.

\section{A brief review}
\label{sec:review}
This section is mainly concerned with self-avoiding walks adsorbed at a 
surface and pulled at an end-point.  This corresponds to the $f=1$ case for stars, 
\emph{i.e.} stars with only one arm.

Vertices in $\mathbb{Z}^d$ have coordinates $(x_1,x_2, \ldots x_d)$, $x_i \in \mathbb{Z}$.
Write $c_n$ for the number of $n$-edge self-avoiding walks starting at the origin.
Then \cite{Hammersley1957}
\begin{equation}
\log d \le \lim_{n\to\infty} \sfrac{1}{n} \log c_n = \log \mu_d \le \log(2d-1)
\end{equation}
where $\mu_d$ is the growth constant of the self-avoiding walk.  
If every vertex of the walk has non-negative $x_d$-coordinate we call the walk a 
\emph{positive walk} and write $c_n^+$ for the number of $n$-edge
positive walks.  Suppose that $c_n^+(v,h)$ is the number
of $n$-edge positive walks with $v+1$ vertices in the hyperplane 
$x_d=0$ and with the $x_d$-coordinate of the last vertex equal to 
$h$.  We say that the walk has $v$ \emph{visits} and the last vertex
has \emph{height} equal to $h$.  Define the partition function
\begin{equation}
C_n^+(a,y) = \sum_{v,h} c_n^+(v,h) a^v y^h.
\end{equation}
Then the free energy
\begin{equation}
\psi(a,y) = \lim_{n\to\infty} n^{-1} \log C_n^+(a,y)
\end{equation}
exists  \cite{Rensburg2013}.  

When $y=1$ (so that there is no force) $\psi(a,1)=\kappa(a)$, the free energy of an 
adsorbing walk \cite{HTW}.  When $a=1$ (so that there is no interaction with the surface)
$\psi(1,y) = \lambda(y)$, the free energy of a 
pulled walk \cite{Beaton2015,IoffeVelenik,IoffeVelenik2010,Rensburg2009}.
There exists a critical value of $a$, $a_c > 1$,
such that $\kappa(a) = \log \mu_d$ when $a \le a_c$ and 
$\kappa(a) > \log \mu_d$ when $a > a_c$, so
that $\kappa(a)$ is singular at $a=a_c > 1$ \cite{HTW,Rensburg1998,Madras}.  Similarly
$\lambda(y)$ is singular at $y=1$ \cite{Beaton2015,IoffeVelenik} and
the walk is in a ballistic phase when $y > 1$.
Moreover, it is known \cite{Rensburg2013} that
\begin{equation}
\psi(a,y) = \max[\kappa(a), \lambda(y)]
\label{eqn:psicondition}
\end{equation}
and, in particular, $\psi(a,y) = \log \mu_d$ when $a \le a_c$
and $y \le 1$.  The case $y < 1$ corresponds to a force directed \emph{towards} the 
surface.  For $a > a_c$ and $y > 1$ there is a phase
boundary in the $(a,y)$-plane along the curve 
given by $\kappa(a) = \lambda(y)$.  This phase transition
is first order \cite{Guttmann2014}.

The free energy of adsorbing walks pulled at their highest vertices was
examined in reference \cite{Rensburg2016b}.  If $c_n(v,h)$ is the number of
positive walks from the origin with length $n$, making $v$ visits, and
with span in the $x_d$-direction (or \emph{height} of the highest vertices)
equal to $h$, and partition function $C_n(a,y)$, then the free energy is
\begin{equation}
\psi (a,y) = \lim_{n\to\infty} \Sfrac{1}{n} \log C_n(a,y)
 = \lim_{n\to\infty} \Sfrac{1}{n} \log C_n^+ (a,y).
\end{equation}
In other words, the free energies of adsorbing walks pulled at their endpoints,
or pulled at their highest vertices, are equal to $\psi(a,y)$.  See
theorem 1 in reference \cite{Rensburg2016b}.

\section{Uniform $3$-stars pulled from an adsorbing surface}

In this section we shall investigate the situation for $3$-stars, attached at
the origin in the $x_3=0$ plane (which is the adsorbing surface) in $\mathbb{Z}^3$ 
at a vertex of unit degree, confined to the half-space $x_3 \ge 0$,  
and pulled, normal to the surface, at another vertex 
of unit degree.  The aim is to compute the (limiting) free energy $\sigma^{(3)}(a,y)$
as a function of the interaction with the surface and the applied force.   Our
strategy is to find matching upper and lower bounds.  Lower bounds are obtained
by strategy bounds; that is, by finding classes of $3$-stars that give a tight
lower bound.  Upper bounds are in general obtained by treating the 
arms of the stars as being independent except when one arm shields
another from the surface (as happens in two dimensions).  

In $d$ dimensions strategy bounds are obtained by 
\begin{enumerate}
\item
working with \emph{bridges} \cite{MadrasSlade} so that the two vertices of degree 
$1$ have largest and smallest $x_d$-coordinates,
\item
working with walks or bridges \emph{unfolded} \cite{HammersleyWelsh} in other
coordinate directions, 
\item
working with \emph{loops} (see section 7.1.4 in reference \cite{Rensburg2015}
and sections 4 and 5 in reference \cite{Guttmann2017}), and
\item
working with walks confined to wedges so that the walks do not interact 
with one another \cite{HammersleyWhittington,Soteros}.
\end{enumerate}

For an $n$-edge self-avoiding walk number the vertices 
$k=0,1, \ldots n$.  We write $x_i(k)$ for the $i$th coordinate of the 
$k$th vertex.  

An $n$-edge self-avoiding walk is a \emph{bridge} if 
$x_d(0) \le x_d(k) < x_d(n)$ for all $1 \le k \le n-1$.  If we write
$b_n$ for the number of $n$-edge bridges then \cite{HammersleyWelsh}
\begin{equation}
b_n \le c_n \le b_n e^{O(\sqrt{n})}.
\end{equation}
We say that 
an $n$-edge self-avoiding walk is \emph{unfolded in the $x_j$-direction}
\cite{HammersleyWelsh}
if $x_j(0) \le x_j(k) \le x_j(n)$ for all $1 \le k \le n-1$.  Notice that a bridge is 
a walk unfolded in the $x_d$-direction.

An $n$-edge self-avoiding walk is a \emph{loop} if $x_d(0)=x_d(n)=0 \le x_d(k)$
for all $1\leq k \leq n-1$ and  is an \emph{unfolded loop} if,
in addition, $x_1(0) \leq x_1(k) \leq x_1(n)$ for
all $1\leq k \leq n-1$.  If the number of loops of length $n$ is denoted by
$\ell_n$,  and the number of unfolded loops by $\ell_n^{\dagger}$ then it is known that 
$\ell_n^{\dagger} \le \ell_n  \leq c_n$ and $\ell^{\dagger}_n = \mu_d^{n+o(n)}$
so that $\lim_{n\to\infty} \sfrac{1}{n} \log \ell_n^{\dagger} =\lim_{n\to\infty} \sfrac{1}{n} \log \ell_n 
=  \log \mu_d$
(see, for example, theorem 7.6 in reference \cite{Rensburg2015}).

%

\subsection{Stars when $d \ge 3$}
\label{sec:pulledstars3}
In this section we concentrate on the three dimensional case, $d=3$.  At the end 
of the section we make some comments about the extension to $d \ge 4$.  

\begin{figure}[t]
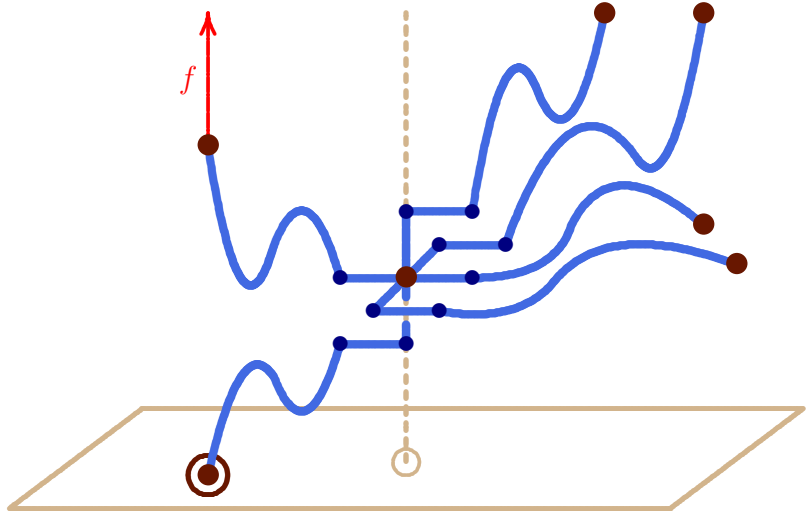

\beginpicture
\setcoordinatesystem units <2.5pt,2.5pt>
\setplotarea x from -85 to 100, y from -5 to 70
\setplotarea x from -60 to 60, y from 0 to 70

\color{Tan}
\setplotsymbol ({\LARGE$\cdot$})

\plot -60 -5 40 -5 60 10 -40 10 -60 -5 /

\setdashes <3pt>
\plot 0 2 0 70 /
\setsolid
\setlinear

\color{Sepia}
\circulararc 360 degrees from -30 3 center at -30 0 

\setplotsymbol ({\scriptsize$\bullet$})
\color{RoyalBlue}

\plot -10 20 0 20 0 23 /   \plot 0 27 0 30 /
\plot 0 30 -10 30 /
\plot 0 30 10 30 /
\plot  0 30 -5 25 5 25 /
\plot 0 30 5 35 15 35 /
\plot 0 30 0 40 10 40 /

\setquadratic
\plot -30 0 -25 15  -20 15 -15 10 -10 20 /
\plot -10 30 -15 40 -20 35 -25 30 -30 50  /

\plot 5 25 15 25 22 29 33 35 50 32 /
\plot 10 30 20 32 25 38 33 44 45 38 /
\plot 15 35 25 52 35 48 40 50 45 70 / 
\plot 10 40 15 60 20 58 25 55 30 70 /

\color{NavyBlue}
\multiput {\Large$\bullet$} at -10 30 -10 20 0 20 0 40 -5 25 5 35 5 25 10 30 15 35 10 40 /

\color{red}
\setplotsymbol ({\large$\cdot$})
\arrow <8pt> [.2,.67] from -30 50 to -30 70 
\put {\large$f$} at -33 60
\color{Sepia}
\multiput {\huge$\bullet$} at -30 0 0 30  -30 50 50 32 45 38 45 70 30 70  /
\color{Tan}
\circulararc 360 degrees from 0 4 center at 0 2 

\color{black}
\normalcolor
\endpicture
\caption{A $6$-star in the half cubic lattice decomposed as two pulled
unfolded bridges to the left of the vertical dividing plane (dashed line), and
four self-avoiding walks to the right of the dividing plane.  The walks
to the right are confined to disjoint infinite wedges in the upper half-space of the cubic lattice.
The geometry about the central vertex of the the star is as illustrated, and arms
can be removed to form $f$-stars with $3\leq f \leq 6$.}
\label{fig:6arm} 
 \end{figure}

The model for $3$-stars is shown in figure \ref{fig:sketch}, and in the cubic lattice
this generalizes to $f$-stars with $3\leq f \leq 6$.  The stars have one vertex
of degree $1$ fixed at the origin (in the adsorbing plane), are confined to the
positive half-lattice, and are pulled by a force $f$ at another vertex of degree
$1$.  The force $f$ is related to the pulling activity $y$ by $y=e^{f/k_BT}$
where $k_B$ is Boltzmann's constant and $T$ is the absolute temperature.
In figure \ref{fig:6arm} a schematic diagram is shown for a $6$-star lattice
polymer pulled by one end-vertex and attached to the origin in the adsorbing
plane at another vertex.

We shall need several Lemmas.

\begin{lemm} [Whittington and Soteros 1991 \cite{WhittingtonSoteros1991}, Soteros 1992 \cite{Soteros}]
If $3 \le f \le 6$ and $d=3$ the free energy of adsorbing $f$-stars is given by
$$\sigma^{(f)}(a,1) = \kappa(a)$$
where $\kappa(a)$ is the free energy of an adsorbing walk \cite{HTW}. \qed
\label{lemma:staradsorb3}
\end{lemm}

In order to prove the next lemma, a short digression on adsorbing and pulled
bridges is needed.  
A bridge of length $n$ and end-point at height $h$ can be
unfolded in the $x_1$-direction to obtain an 
\emph{unfolded bridge} \cite{HammersleyWelsh} of length $n$ and height $h$.  If
$b_n(h)$ is the number of bridges from the origin
of length $n$ and endpoint at height $h$, and $b_n^\dagger(h)$
is the number of unfolded bridges of length $n$ and height $h$ then
$b_n^\dagger(h) \leq b_n (h) \leq e^{o(n)}b_n^\dagger(h)$.
By theorems 1 and 3 in reference \cite{Rensburg2016a} this, in particular, 
shows that,  if the partition function of a pulled unfolded bridge is
$B_n^\dagger(y) = \sum_h b_n^\dagger(h)\, y^h$, then
\begin{equation}
 \lim_{n\to\infty} \sfrac{1}{n} \log B_n^\dagger(y) = \lambda(y).
\label{eqnB}   
\end{equation}
This result is now used to prove lemma \ref{lemma:starpull3}.

\begin{lemm}
When there is no interaction with the surface ($a=1$) 
or a repulsive interaction with the surface ($a < 1$), the free
energy of stars when $3 \le f \le 6$ and $d=3$ is given by
$$\sigma^{(f)}(a,y) = \Sfrac{2}{f}\, \lambda(y) + \Sfrac{f-2}{f} \, \log \mu_3.$$
\label{lemma:starpull3}
\end{lemm}
\Pr
First consider the case $a=1$ and write $n=fm$.  A pulled uniform $f$-star of length
$fm$ can be decomposed into a pulled walk from the origin
of length $2m$, and $f-2$ remaining branches or arms each 
of length $m$.  This shows that
\[ \limsup_{m\to\infty} \Sfrac{1}{fm} \log S_{fm}^{(f)}(1,y)
\leq \Sfrac{2}{f} \lambda(y) + \Sfrac{f-2}{f} \log \mu_3 . \]
A lower bound is found by creating a pulled $f$-star from
pulled unfolded walks and bridges.

A pulled $f$-star can be decomposed as shown in figure 
\ref{fig:6arm} into two pulled unfolded bridges and $f-2$ self-avoiding
walks confined in disjoint infinite wedges in the cubic lattice
that avoid the $x_3=0$ plane.  In figure
\ref{fig:6arm} the schematic is for a $6$-star; see for
example section 11.5 in reference \cite{Rensburg2015} for more
general results.  Notice the geometry of the arms of the star
close to the central vertex as shown in figure \ref{fig:6arm}.
Since the growth constant for self-avoiding walks 
of length $m$ in an infinite wedge is $\mu_3$ \cite{HammersleyWhittington}, the result 
is that
\[ \liminf_{m\to\infty} \Sfrac{1}{fm}  \log S_{fm}^{(f)}(1,y)
\geq \Sfrac{2}{f}\lambda(y) + \Sfrac{f-2}{f} \log \mu_3 . \]

It remains to consider the case that $a<1$.
A lower bound on $S_{fm}^{(f)}(0,y)$ is obtained as in figure
\ref{fig:6arm}, by decomposing the star into bridges and walks,
but now with the only difference that there are no visits to the
adsorbing plane. 
The arms on the left are bridges, while those on the right are
walks in infinite wedges in the cubic lattice.  By removing some of these 
arms,  stars with $3\leq f \leq 6$ can be created. This shows that
\[ \liminf_{n\to\infty} \Sfrac{1}{fm} \log S_{fm}^{(f)}(0,y)
 \geq \Sfrac{2}{f}\lambda(y) + \Sfrac{f-2}{f} \log \mu_3 . \]
By monotonicity $S_{fm}^{(f)}(0,y) \leq S_{fm}^{(f)} (a,y)
\leq S_{fm}^{(f)} (1,y)$ if $0 < a < 1$.  This completes the proof.
\qed









If $b_n^\dagger(v,h)$ is the number of unfolded bridges with $v$ visits
and height $h$, then the unfolded adsorbing bridge partition function is given by
\begin{equation}
B_n^\dagger(a,y) = \sum_{v=0}^n \sum_{h=0}^n b_n^\dagger (v,h)\, a^v y^h .
\end{equation}
It is known that $\lim_{n\to\infty} \Sfrac{1}{n} \log B_n^\dagger (1,y) = \lambda (y)$
(see equation (22) in reference \cite{Rensburg2017}).  Since $B_n^\dagger(0,y) 
= B_{n-1}^\dagger(1,y)$, it follows that
\begin{equation}
\lim_{n\to\infty} \Sfrac{1}{n} \log B_n^\dagger (a,y) = \lambda (y),
\quad\hbox{for all $0\leq a \leq 1$},
\label{eqn12}   
\end{equation}
by monotonicity.

\subsection{Lower bounds on the free energy for $3$-stars in 3 dimensions}
\begin{figure}[t]
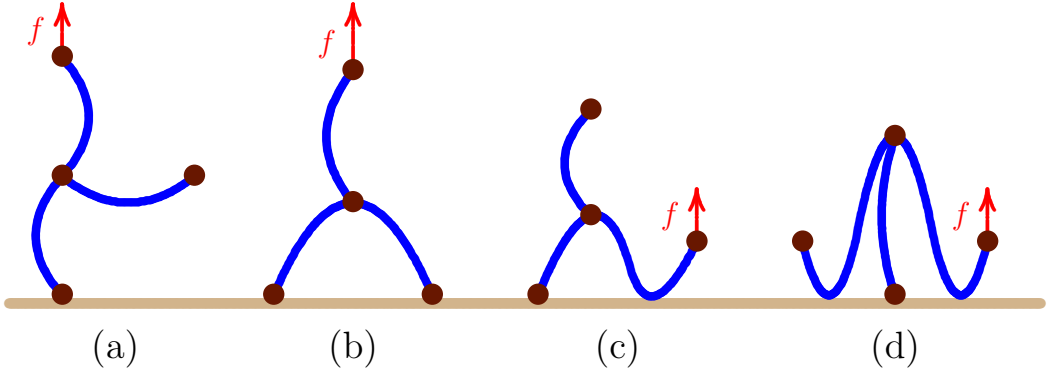

\beginpicture
\setcoordinatesystem units <1.0pt,1.0pt>
\setplotarea x from -140 to 150, y from -20 to 110
\setplotarea x from -100 to 250, y from 0 to 90

\color{Tan}
\setplotsymbol ({$\bullet$})
\plot -120 -3 270 -3 /

\setplotsymbol ({\scriptsize$\bullet$})
\color{Blue}
\setquadratic
\plot -100 0 -110 22.5 -100 45 -90 67.5 -100 90 /
\plot -100 45 -75 35 -50 45 /

\plot -20 0 10 35 40 0 /
\plot 10 35 0 60 10 85 /

\plot 80 0 100 30 115 10 125 0 140 20 /
\plot 100 30 90 50 100 70 /

\plot 180 20 190 0 200 20 215 60 230 20 240 0 250 20  /
\plot 215 0 210 30 215 60 /

\color{red}
\setplotsymbol ({\large$\cdot$})
\arrow <8pt> [.2,.67] from -100 90 to -100 110 
\put {\large$f$} at -110 100
\arrow <8pt> [.2,.67] from 10 85 to 10 110 
\put {\large$f$} at 0 95 
\arrow <8pt> [.2,.67] from 140 20 to 140 40 
\put {\large$f$} at 130 30 
\arrow <8pt> [.2,.67] from 250 20 to 250 40 
\put {\large$f$} at 240 30 


\color{Sepia}
\multiput {\huge$\bullet$} at 
-100 0 -100 90 
-50 45  -100 45
-20 0 40 0 10 35 10 85
80 0 100 30  140 20 100 70 
180 20 250 20 215 60 215 0 /

\color{black}
\put {\Large(a)} at -80 -20 
\put {\Large(b)} at 10 -20 
\put {\Large(c)} at 110 -20 
\put {\Large(d)} at 215 -20 
\normalcolor
\setlinear
\endpicture
\caption{The cases for finding a strategy lower bound on the free energy of
pulled and adsorbing $3$-stars in the cubic lattice.}
\label{fig:cases}  
\end{figure}

In lemma \ref{lemma:starpull3} the free energy of adsorbing $f$-stars
is given for $y>0$ and $0 \leq a \leq 1$.  We next examine lower
bounds on the free energy, particularly when $a \ge 1$ and $y \ge1$.
The lower bounds are constructed using \emph{strategy arguments}.  
The cases that need to be considered are shown in figure \ref{fig:cases} and are, from the
left, a case where only one of the arms of the star has vertices in the adsorbing
plane, two cases where two arms have vertices in the adsorbing plane, 
and then one case where all three arms have vertices in the adsorbing plane.

In this section assume that $n=fm$ with $f=3$, and all limits $n\to\infty$ 
are taken through multiples of $3$, namely $n=3m$ and $m\to\infty$.

\begin{lemm} 
If $a \geq 1$, then
$\displaystyle
\liminf_{n\to\infty} \Sfrac{1}{n} \log S_{n}^{(3)}(a,y) \geq 
\Sfrac{1}{3}[2  \lambda(y) +  \log \mu_3]  $.
\label{lemma3}
\end{lemm}
\Pr
This bound is a consequence of case (a) in figure \ref{fig:cases}.  By monotonicity
$S_n^{(3)} (a,y) \ge S_n^{(3)} (1,y)$ for $a \ge 1$ so that
$$\liminf_{n\to\infty} \Sfrac{1}{n} \log S_{n}^{(3)}(a,y) \geq \sigma^{(3)}(1,y).$$  Then using Lemma \ref{lemma:starpull3}
with $a=1$ and $f=3$ gives the required result.
\qed

\begin{lemm}
If $a \ge 1$, $y \ge 1$ and $d=3$ then
$\displaystyle
\liminf_{n\to\infty} \Sfrac{1}{n} \log S_{n}^{(3)}(a,y) \geq 
\Sfrac{1}{3} [2\kappa(a) +  \lambda(y)]. $
\label{lemma4}
\end{lemm}
\Pr
This comes from a special version of case (b) in figure \ref{fig:cases}.  Write $n=3m$.  Consider a 
loop with $m$ edges unfolded in the $x_1$ and $x_2$-directions so that it is confined to the 
octant $x_1 \le 0$, $x_2 >0$ and $x_3\ge 0$ except that the loop starts 
at the origin and the first edge steps along the positive $x_2$-axis.  Similarly consider a 
loop with $m$ edges unfolded in the $x_1$ and $x_2$-directions so that it is confined to the 
octant $x_1 \le 0$, $x_2 < 0$ and $x_3\ge 0$ except that the loop starts 
at the origin and the first edge steps along the negative $x_2$-axis.  These two loops
together form a loop with $2m$ edges that avoids the half-space $x_1>0$.  Their combined 
contribution to the free energy is $\sfrac{2}{3} \kappa(a)$ \cite{HTW}.  Now add a bridge
with $m$ edges, unfolded in the 
$x_1$-direction, starting at the origin, having its first edge along the positive
$x_3$-axis and its second edge
from $(0,0,1)$ to $(1,0,1)$, so that it otherwise avoids the half-space $x_1 \le 0$. 
This unfolded bridge
contributes $\sfrac{1}{3}\lambda(y)$ to the total free energy and adding these terms
completes the proof (by equation \Ref{eqn12}, since the bridge has no visits).
\qed

\begin{lemm}
If $a \ge 1$, $y \ge 1$ and $d=3$ then
$\displaystyle
\liminf_{n\to\infty} \Sfrac{1}{n} \log S_{n}^{(3)}(a,y) \geq 
\kappa(a).$
\label{lemma5}
\end{lemm}
\Pr  
This bound is a consequence of case (d) in figure \ref{fig:cases}.  By monotonicity
$S_n^{(3)} (a,y) \ge S_n^{(3)} (a,1)$ for $y \ge 1$ so that
$$\liminf_{n\to\infty} \Sfrac{1}{n} \log S_{n}^{(3)}(a,y) \geq \sigma^{(3)}(a,1).$$  Then using Lemma \ref{lemma:staradsorb3}
with $f=3$ gives the required result.
\qed

These three lower bounds, taken together, prove the following result:

\begin{lemm}
When $a \ge 1$, $y \ge 1$ and $d=3$
$$\liminf_{n\to\infty} \Sfrac{1}{n} \log S_{n}^{(3)}(a,y) \geq 
\max\left[ \Sfrac{1}{3}[2 \lambda(y) +  \log \mu_3],   \Sfrac{1}{3}[2 \kappa(a) +  \lambda(y)], \kappa(a)
\right].$$
\label{lemma:lowerbound}
\end{lemm}
Considering case (c) in figure \ref{fig:cases} does not give a useful bound.


\subsection{Upper bounds on the free energy for $3$-stars in 3 dimensions}

To obtain upper bounds corresponding to the lower bounds derived in the last
section we use a case analysis, as sketched in  figure \ref{fig:cases}.  We again consider in turn 
the cases where exactly one, two or three of the branches have vertices in
the plane $x_3=0$.  As before, $n=fm$ with $f=3$, and all limits $n\to\infty$ 
are taken through multiples of $3$, namely $n=3m$ and $m\to\infty$.

\begin{lemm}
When $f=3$ and $d=3$, if exactly one branch has vertices in $x_3=0$ then
the contribution to the free energy is at most
$\sfrac{1}{3}
\left( \max [\kappa(a),\lambda(y)] + \lambda(y) + \log \mu_3 \right).$
\end{lemm}
\Pr
This case corresponds to case (a) in figure
\ref{fig:cases}.
We obtain an upper bound by considering the two branches from the 
origin to the vertex at which the force is applied as independent of the 
third branch.   Of the first two branches only one can have 
vertices in $x_3=0$ so together they contribute 
$\sfrac{1}{3}\left( \max[\kappa(a), \lambda(y)] + \lambda(y) \right)$ to the free energy and the 
third branch contributes $\sfrac{1}{3}\log \mu_3$ since it is not subject to 
the force and has no vertices in $x_3=0$.  This proves the lemma.
\qed

When two branches have vertices in $x_3=0$ (see figure \ref{fig:cases}(b) and (c))
we have to consider two 
subcases: 
\begin{enumerate}
\item
when the vertex at which the force is applied \emph{is not} in a branch with 
vertices in $x_3=0$, and
\item
when the vertex at which the force is applied \emph{is}  in a branch with 
vertices in $x_3=0$.
\end{enumerate}

\begin{lemm}
When $f=3$ and $d=3$, if exactly two branches have vertices in $x_3=0$ and the vertex
at which  the force is applied is not in a branch with vertices in $x_3=0$ then
the contribution to the free energy is at most
$\sfrac{1}{3} \left( \max [2\kappa(a),2\lambda(\sqrt{y})] + \lambda(y)  \right).$
\end{lemm}
\Pr
This corresponds to case (b) in figure \ref{fig:cases}.  Since the branch at which 
the force is applied does not have vertices in $x_3=0$ it contributes 
$\sfrac{1}{3}\lambda(y)$ to the free energy.  The other two 
branches form a loop (which is being pulled) together with vertices in the 
surface $x_3=0$.  The free energy of a pulled loop is $\lambda(\sqrt{y})$
(see reference \cite{Guttmann2017}), 
and so this gives a free energy bounded above by a linear 
combination of $\sfrac{2}{3}\lambda(\sqrt{y})$ and $\sfrac{2}{3}\kappa(a)$.  
This linear combination is bounded above by 
$\sfrac{2}{3}\max[\lambda(\sqrt{y}),\kappa(a)]$.  Adding these
terms gives the required upper bound.
\qed

\begin{lemm}
When $f=3$ and $d=3$, if exactly two branches have vertices in $x_3=0$ and the vertex
at which  the force is applied is in a branch with vertices in $x_3=0$ then
the contribution to the free energy is at most
$\sfrac{1}{3}\left( \max [\kappa(a),\lambda(y)] + \kappa(a)  +  \log \mu_3\right).$
\end{lemm}
\Pr
This corresponds to case (c) in figure \ref{fig:cases}.    Since the branch at which the force is 
applied has at least one vertex in the plane $x_3=0$ the free energy contribution from this branch 
is at most $\sfrac{1}{3}\max[\kappa(a),\lambda(y)]$.  The branch containing the origin contributes
at most $\sfrac{1}{3}\kappa(a)$.  The remaining branch is not subject to a force and has no
vertices in $x_3=0$ and so contributes $\sfrac{1}{3} \log \mu_3$.  Adding these three terms
gives the required upper bound.
\qed

The final case to be considered is when all three branches have vertices in
$x_3=0$.

\begin{lemm}
When $f=3$ and $d=3$, if all three  branches have vertices in $x_3=0$ then
the contribution to the free energy is at most
$\sfrac{1}{3}\left(  2 \kappa(a) +\max [\kappa(a),\lambda(y)] \right).$
\end{lemm}
\Pr  This corresponds to case (d) in figure \ref{fig:cases}.  Since all three branches have vertices 
in $x_3=0$, in particular the branch at which the force is applied has 
vertices in $x_3=0$.  The maximum contribution to the free energy 
from this branch is $\sfrac{1}{3}\max[\kappa(a), \lambda(y)]$.  The 
other two branches are not under tension so they each contribute a 
maximum of $\sfrac{1}{3}\kappa(a)$.  Adding these three terms
gives the required bound.
\qed

Since $a \ge 1$ and $y \ge 1$, $\kappa(a) \ge \log \mu_3$ and $\lambda(y) 
\ge \log \mu_3$.  In addition, since $\lambda(y)$ is a convex function 
of $\log y$, 
\begin{equation}
\lambda(\sqrt{y} ) \le \sfrac{1}{2}(\lambda(y) + \log \mu_3).
\end{equation}
Using these results and the results of the last three lemmas gives 
the following result:

\begin{lemm}
When $a\ge 1$, $y\ge 1$, $f=3$ and $d=3$ then
$$\limsup_{n\to\infty} \Sfrac{1}{n} \log S_{n}^{(3)}(a,y) \le \max[ \kappa(a),
 \sfrac{1}{3}( 2\lambda(y)+\log \mu_3),\sfrac{1}{3}(\lambda(y) + 2 \kappa(a) )].$$
\end{lemm}

These upper bounds exactly match the lower bounds in lemma
\ref{lemma:lowerbound} giving the result

\begin{theo}
When $a \ge 1$, $y \ge 1$, $f=3$ and $d=3$ then
$$\lim_{n\to\infty} \Sfrac{1}{n} \log S_{n}^{(3)}(a,y) = \sigma^{(3)}(a,y)  
= \max[ \kappa(a),\sfrac{1}{3} (2\lambda(y)+\log \mu_3) ,
\sfrac{1}{3} (\lambda(y) + 2 \kappa(a))].$$
\end{theo}

The extension of these results from $d=3$ to $d \ge 3$ produces no new complications.
However, the extension to $d=2$ is not trivial because branches can shield other 
branches from the surface.  Some of the free energy bounds when $d=2$ will be 
different from those for $d \ge 3$.


\subsection{3-stars with the force directed towards the surface}

In this section we examine the situation when the force is directed 
towards the surface so that $y < 1$.  Our primary result is the following 
theorem:

\begin{theo}
When $d=3$, $f=3$ and $y \le 1$ the free energy is equal to 
$\kappa(a)$.  In particular, when $y \le 1$ and $a \le a_c$
the free energy is $\log \mu_3$.
\end{theo}
\Pr
When $y=1$ the free energy is equal to $\sigma^{(3)}(a,1) = \kappa(a)$.
By monotonicity 
\begin{equation}
\limsup_{n\to\infty} \Sfrac{1}{n} \log S_{n}^{(3)}(a,y) \le \kappa(a)
\end{equation}
for all $y < 1$.  To get a lower bound  we look at the situation when 
$y=0$, so that the vertex at which the force is applied is in
$x_3=0$, and use monotonicity in $y$.  Consider stars with two vertices of degree
1 and the vertex of degree 3 in $x_3=0$.  (The force is applied at one of these vertices
of degree 1.)  The star can be thought of as two loops and one positive
walk starting at the origin.  By a construction similar to that used in the proof
of lemma \ref{lemma4} we see that $\kappa(a)$ is a lower bound on the free energy,
completing the proof.
\qed


\section{The phase diagram}
\label{sec:phasediagram}

The results given above give a considerable amount of information about the phase 
diagram in the $(a,y)$-plane for 3-stars in the simple cubic lattice.  

If $a < a_c$ and $y < 1$ the free energy is $\log \mu_3$ and we have a 
\emph{free phase}.  If $a > a_c$ and $\lambda(y) < \kappa(a)$ the free 
energy is $\kappa(a)$ and we have an \emph{adsorbed phase}.  If 
$y > 1$ and $\lambda(y) > 2 \kappa(a) - \log \mu_3$ the free energy is 
$\frac{1}{3}(2 \lambda(y) + \log \mu_3)$ and we have a \emph{ballistic phase}.
If $2\kappa(a) - \log \mu_3 > \lambda(y) > \kappa(a)$ then the
free energy is $\frac{1}{3}(2\kappa(a) + \lambda(y))$ and we have a 
\emph{mixed phase}.   Note that the free energy depends only on $y$ 
in the ballistic phase, only on $a$ in the adsorbed phase but
on both $a$ and $y$ in the mixed phase.

The phase boundary between the free phase and the adsorbed phase 
is given by $a=a_c$, the adsorption critical point for self-avoiding walks.
The phase boundary between the free phase and the ballistic phase
is given by $y=1$, the ballistic critical point for pulled walks.  Between
the adsorbed phase and the mixed phase there is a phase boundary
$y=y_1(a)$ (for $a\geq a_c$) given by the solution of
\begin{equation}
\lambda(y) = \kappa(a), \qquad\hbox{for $a\geq a_c$},
\end{equation}
and since $\lambda(y)$ is monotonic increasing when $y > 1$, this shows 
that $y_1(a) = \lambda^{-1}(\kappa(a))$.

The phase boundary $y=y_2(a)$ separating the mixed and ballistic phases
is similarly given by the solution of
\begin{equation}
\lambda(y) = 2\kappa(a) - \log \mu_3, \qquad\hbox{for $a\geq a_c$},
\end{equation}
and this is given by $y_2(a) = \lambda^{-1}(2\kappa(a) - \log \mu_3)$.

\def\axes#1#2#3#4#5#6#7{
\setplotarea x from #7 to #5, y from #2 to #6
\setplotarea x from #1 to #5, y from #2 to #6
\axis left shiftedto x=#3 /
\axis bottom shiftedto y=#4 /
\put {\footnotesize$\bullet$} at #3 #4
}

\begin{figure}[t]
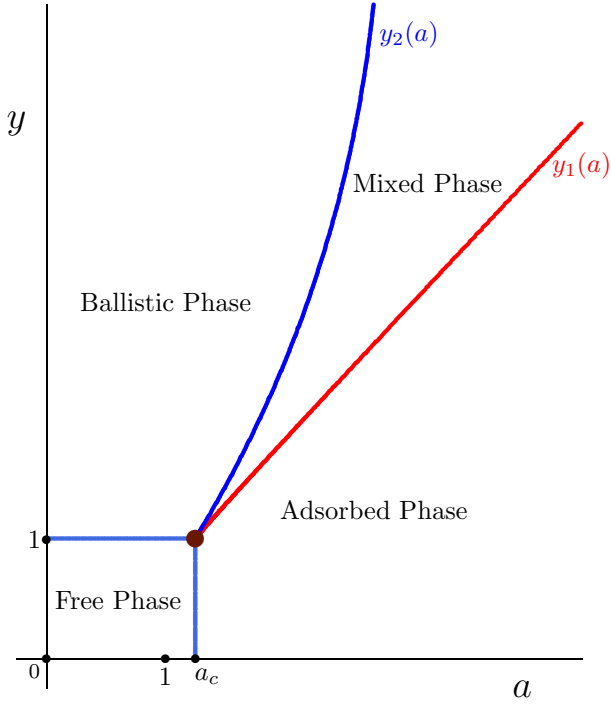

\begin{center}
\beginpicture

\color{black}
\setcoordinatesystem units <2.25pt,2.25pt>

\setplotsymbol ({\Large$\cdot$})
\color{RoyalBlue}
\plot 0 20 25 20 25 0 /

\setquadratic
\color{Red}
\plot 25 20 45 42 90 90 /
\put {$y_1(a)$} at 90 83

\color{Blue}
\plot 25 20 45 63 55 110 /
\put {$y_2(a)$} at 61 105

\setlinear
\color{black}
\axes{-5}{-5}{0}{0}{90}{110}{-40}
\put {\footnotesize$0$} at -2 -2
\put {\footnotesize$\bullet$} at 25 0
\put {$a_c$} at 27 -3
\put {\footnotesize$\bullet$} at 20 0
\put {$1$} at 20 -3
\put {\footnotesize$\bullet$} at 0 20
\put {$1$} at -2 20

\color{Sepia}
\put {\LARGE$\bullet$} at 25 20 

\color{black}
\put {Free Phase} at 12 10 
\put {Adsorbed Phase} at 55 25 
\put {Mixed Phase} at 64 80
\put {Ballistic Phase} at 20 60 

\put {\Large$a$} at 80 -5
\put {\Large$y$} at -5 90

\color{black}
\normalcolor

\endpicture
\end{center}
\caption{The phase diagram in the $(a,y)$-plane for 3-stars on the simple 
cubic lattice.  There are four phases: a free phase, a ballistic phase, an adsorbed
phase and a mixed phase.  In the mixed phase the free energy depends on
both $a$ and $y$.}
\label{fig:phasediagram}   
\end{figure}

The phase diagram is shown schematically in figure \ref{fig:phasediagram}.  Notice
that asymptotically $\lambda(y) \sim \log y$ \cite{Rensburg2013} and $\kappa(a) \sim \log a
+ \log \mu_2$ \cite{Rensburg2013,RychlewskiJSP}.  This shows that, for $a$ large,
\begin{eqnarray}
y_1(a) \sim & \mu_2\, a, &\quad\hbox{between the adsorbed and mixed phases};  \\
y_2(a) \sim & \L \sfrac{\mu_2^2}{\mu_3}\R  a^2, 
&\quad\hbox{between the mixed and ballistic phases.}
\end{eqnarray}
Since $\sigma^{(3)}(a,y)$ is a constant function of $a$ for fixed $y$ in the ballistic phase, and
a constant function of $y$ for fixed $a$ in the adsorbed phase, both the phase boundaries
$y_1(a)$ and $y_2(a)$ correspond to first order phase transitions in the model.  To see this,
we observe the following:
\begin{enumerate}
\item
Since $\kappa(a)$ and $\lambda(y)$ are convex functions of $\log a$
and $\log y$ respectively \cite{HTW,Rensburg2013}, they are differentiable 
almost everywhere and have left and right 
derivatives everywhere.
\item
$y_1(a) > 1$ for all $a > a_c$, the free energy is independent of $y$ in the 
adsorbed phase and is equal to $\sfrac{1}{3}(2\kappa(a) + \lambda(y))$ in the mixed phase
so its left partial derivative with respect to $y$ is positive throughout this phase.  
Therefore the left partial derivative of the free energy with respect to $y$
has a jump discontinuity on the phase boundary $y=y_1(a)$.
\item
In a similar way, 
$y_2(a) > 1$ for all $a > a_c$, the free energy is independent of $a$ in the 
ballistic phase and is equal to $\sfrac{1}{3}(2\kappa(a) + \lambda(y))$ in the mixed phase
so its left partial derivative with respect to $a$ is positive throughout this phase.  
Therefore the left partial derivative with respect to $a$
of the free energy has a jump discontinuity on the phase boundary $y=y_2(a)$.
\end{enumerate}

In order to switch to the force-temperature plane we write $y=\exp[f/k_BT]$ and
$a=\exp[1/k_BT]$ where $f$ is the applied force, $T$ is the absolute temperature 
and $k_B$ is Boltzmann's constant.  At low temperature the phase boundary between the 
adsorbed and mixed phases behaves as
\begin{equation}
f = f_1(T) \simeq 1 + ( \log\mu_2) k_BT
\end{equation}
and the phase boundary between the ballistic and mixed phases behaves as
\begin{equation}
f  = f_2(T) \simeq 2 + (2\log \mu_2 - \log \mu_3) k_BT.
\end{equation}
Since $\mu_2^2 > \mu_3$ both phase boundaries are re-entrant and 
$\lim_{T\to 0} \sfrac{d}{dT} f_j(T) > 0$ for $j=1$ and $2$.  This is associated with
a loss of entropy in going from the adsorbed phase to the mixed phase and 
from the mixed phase to the ballistic phase.

\section{Discussion}
\label{sec:discussion}

We have investigated rigorously the behaviour of uniform 3-stars
on the simple cubic lattice, with a vertex of degree 1 in the surface at 
which they can adsorb, and pulled at another vertex of degree 1.  We
have established the dependence of the free energy on the 
parameters $a$ and $y$, related to the interaction strength with 
the surface and the applied force, and the resulting phase diagram
in the $(a,y)$-plane.  There are four phases: a free phase where the 
free energy is constant, a ballistic phase where the free energy depends
only on $y$, an adsorbed phase where the free energy depends only
on $a$, and a mixed phase where it depends on both $a$ and $y$.
Mixed phases have been seen in a 
two dimensional directed model of adsorbing polygons
subject to a force \cite{Beaton2017} and have been predicted numerically for a 
similar model for self-avoiding polygons in two dimensions \cite{Guttmann2017}.
They can also occur when an adsorbed self-avoiding walk is pulled at an
interior vertex \cite{Rensburg2017}.

We believe that this is the first case in which a phase diagram 
with a mixed phase has been rigorously established in complete detail for a 
non-directed self-avoiding walk model of a polymer.

\section*{Acknowledgement}
This research was partially supported by  Discovery Grants from NSERC of Canada and
by the Leverhulme Trust Research Programme Grant No. RP2013-K-009. 
SGW would like to acknowledge 
the hospitality of University of Bristol where part of this work was carried out.

\end{document}